\documentclass[08pt,twocolumn,preprintnumbers,amsmath,amssymb,footinbib,superscriptaddress,prl]{revtex4-2}
\usepackage{sistyle}
\usepackage{ulem} 
\usepackage{verbatim}
\usepackage[dvipsnames]{xcolor}
\usepackage{graphicx}
\usepackage{dcolumn}
\usepackage{bm}

\usepackage{braket}
\newcommand{\D}{\operatorname{d\!}}

\begin{document}


\title{Coherent expansion of the motional state of a massive nanoparticle beyond its linear dimensions}

\author{R.~Muffato}
\affiliation{School of Physics and Astronomy, University of Southampton, SO17 1BJ, Southampton, England, UK}
\affiliation{Department of Physics, Pontifical Catholic University of Rio de Janeiro, Brazil}
\author{T.S.~Georgescu}
\affiliation{School of Physics and Astronomy, University of Southampton, SO17 1BJ, Southampton, England, UK}
\author{M.~Carlesso}
\affiliation{Department of Physics, University of Trieste, Strada Costiera 11, 34151 Trieste, Italy}
\affiliation{Istituto Nazionale di Fisica Nucleare, Trieste Section, Via Valerio 2, 34127 Trieste, Italy}

\author{M.~Paternostro}
\affiliation{Universit\`a degli Studi di Palermo, Dipartimento di Fisica e Chimica - Emilio Segr\`e, via Archirafi 36, I-90123 Palermo, Italy}
\affiliation{Centre for Quantum Materials and Technologies, School of Mathematics and Physics, Queen's University Belfast, BT7 1NN, United Kingdom}
\author{H.~Ulbricht}
\email[Correspondence email address: ]{H.Ulbricht@soton.ac.uk;}
\affiliation{School of Physics and Astronomy, University of Southampton, SO17 1BJ, Southampton, England, UK}

\date{\today}

\begin{abstract}
\noindent
Quantum mechanics predicts that massive particles exhibit wave-like behavior. Matterwave interferometry has validated such predictions through ground-breaking experiments involving microscopic systems like atoms and molecules. The wavefunction of such systems coherently extends over a distance much larger than their size, an achievement that is incredibly challenging for  massive and more complex objects. Yet, reaching similar level of coherent expansion will enable tests of fundamental physics at the genuinely  macroscopic scale, as well as the development of quantum sensing apparata of great sensitivity. Here, we report on experimentally achieving an unprecedented degree of position expansion in a massive levitated optomechanical system through frequency modulation of the trapping potential. By starting with a pre-cooled state of motion and employing a train of sudden pulses yet of mild modulation depth, we surpass previously attained values of position expansion in this class of systems to reach expansion lengths that exceed the physical dimensions of the trapped nanoparticle.
\end{abstract}

\maketitle

Achieving large quantum coherence is a crucial step for revealing the wave-like nature of quantum states. Delocalised coherent wavepackets represent valuable resources for quantum sensing applications~\cite{degen2017quantum}, quantum computation with continuous variables~\cite{lloyd1999}, and to perform tests of fundamental physics addressing the boundaries between classical and quantum mechanics~\cite{Bassi2013,Carlesso2022,Belenchia2022,Hornberger2012}.

Time-of-flight techniques~\cite{neumeier2024fast,bonvin2024state}, which involves turning off the optical trap, remain the gold standard to achieve large degrees of coherent inflation of the wavefunction of massive nanoparticles. However, such technique is limited by  gravitational effects, thus requiring microgravity conditions~\cite{kaltenbaek2023research}, and the need for careful  time-gating  of control operations, such as release and recapture mechanisms, so as to allow the nanoparticle's initial wavepacket to evolve and expand. Moreover, the natural evolution of the wavefunction for massive particles is very slow, as its speed scales with the inverse of the mass. A very promising alternative approach involves partially lowering the height of a confining potential well to facilitate state expansion of a particle, while maintaining a sufficient degree of spatial confinement. By dynamically adjusting the well depth, with accurate timing, the motional state of the quasi-trapped particle can be expanded by many orders of magnitude, and much faster than the inverse-of-mass rate typical of nanoparticle-based interferometers. 

Here, we make a significant step forward towards the achievement of spatially extended quantum-coherent states of motion by using a simple yet very effective control strategy: we drive the dynamics of an initially cooled mechanical oscillator through a sequence of  pulses, achieving an expanded state -- in phase space -- associated with a degree of motional delocalisation of $28.4$\,dB. {Moreover, the coherence in the engineered motional state grows in time up until when the effects of the noise due to the laser become too large. 
Our experimental results underpin 
-- through technical improvements to our platform -- 
the possibility to engineer valuable resources, based on the physics of levitated massive nanoparticles, towards the achievement of a genuinely quantum-enhanced regime of sensing.}

\begin{figure*}
\includegraphics{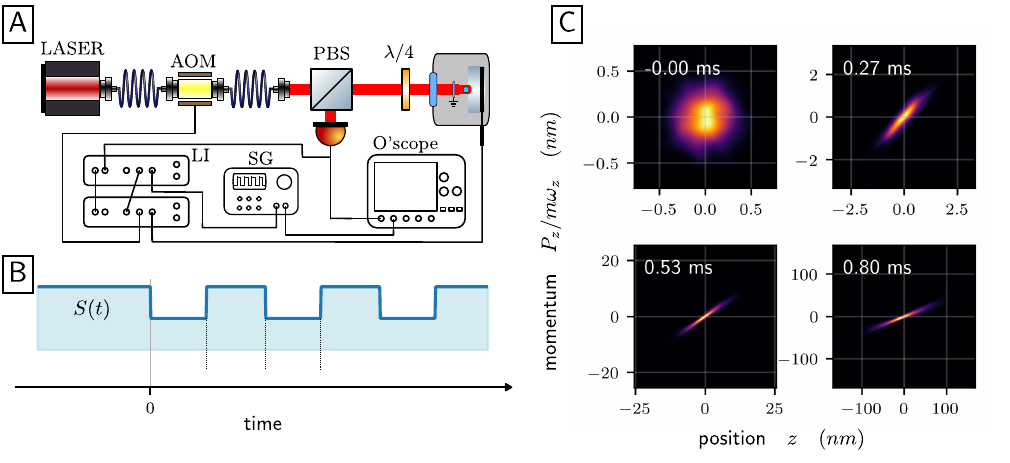}
\caption{\label{fig:wide_panel}{\bf Experimental setup for pulse-driven coherent wakepacket expansion} {\it Panel A}: Schematic of experimental setup. A silica nanoparticle is trapped by an optical tweezers in vacuum. An
acusto-optical modulator (AOM) controlled by a signal generator (SG) and lock-in amplifier (LI) modulates the power of the laser, which serves as the mechanism to modulate the intensity of the driving laser and thus control the trapping potential of the particle. The oscilloscope (O'scope), triggered by a synchronous signal from the SG, records
the signal from the particle. The polarising beam splitter (PBS) and $\lambda/4$ waveplate routes the particle's movement signal towards a photo-detector. The particle displacement is detected to implement parametric and electric cooling mechanisms to lower the energy of the translational degrees of freedom of the system in three directions (only one direction shown in the figure). {\it Panel B}: Time sequence of laser modulation $S(t)$ for expanding the mechanical state with timings for lowering the power $\tau_{low}$ and rising it back up $\tau_{high}$. {\it Panel C}: Phase-space distribution of position and momentum for the nanoparticle. From top-left subpanel we show the evolution of an isotropic initial state at \SI{4.2}{mK} within a \SI{0.8}{ms} timeframe.}
\end{figure*}

\noindent
{\it Experimental setting} -- The experimental setup is illustrated in Fig.~\ref{fig:wide_panel}A. We use a parabolic mirror to focus linearly $x$-polarized light at the wavelength of \SI{1550}{nm} to a diffraction-limited spot of the diameter of $1\,\mu$m. A nearly spherical silica particle with a diameter of \SI{200}{nm} and a mass of $m={7.5\times 10^{-18}}\,\text{kg}$, calculated using the density of silica $\rho=1800$\,$\text{kg}/\text{m}^3$, is trapped at the focal point of the light after being launched from a spray. An acusto-optical modulator is used to modulate the laser intensity and control the trapping potential dynamically. We monitor the centre-of-mass motion of the trapped particle by detecting a small fraction of Rayleigh back-scattered light from the trapping field with a single photodiode detector \cite{vovrosh2017parametric}. The vacuum chamber is evacuated to a pressure of \SI{5}{mbar}, ensuring effective thermal coupling to the residual gas and resulting in the appearance of distinct spectral peaks corresponding to the translational degrees of freedom with frequencies of {$\omega_{z}/2\pi=\SI{77.6}{kHz}$}, {$\omega_{y}/2\pi=\SI{106.4}{kHz}$} and {$\omega_{x}/2\pi=\SI{137}{kHz}$}. 

By fitting a Lorentzian profile to the spectrum of each harmonic trap frequency and applying the equipartition theorem at an assumed temperature of \SI{300}{K}, we calibrate the system and subsequently calculate the position variance of the $z$ translational mode as $\sigma_{300K}^{2}=(2.30 \pm 0.03) \times 10^{-15} \, \text{m}^2$. The detected signal of particle's displacement is used to cool its three translational degrees of freedom, while the pressure is brought down to \SI{3e-7}{mbar}.  The transversal $x$ and $y$ modes are cooled by parametric feedback all the way down to a level ensuring that the translational degrees of freedom are mutually decoupled. We can thus focus on one mode only and treat the problem as quasi-one-dimensional for the $z$ axis. Such mode  is cooled electrically by cold damping using an electrode actuating on the charged particle with a uniform electric field proportional to the particle's $z$ velocity. The cooled state has a position variance $\sigma_\text{cold}^{2}=(3.21 \pm 0.11) \times 10^{-20} \, \text{m}^2$, which corresponds to an effective centre-of-mass temperature of 
\begin{equation}
T_\text{eff} =\SI{300}{K}\times\frac{\sigma^{2}_\text{cold}}{\sigma^{2}_{300{\text K}}}=(4.18 \pm 0.15) \,\text{mK},
\end{equation}
and occupation number $\bar n=(e^{\hbar\omega_z/k_B T_\text{eff}}-1)^{-1}=(1154.36 \pm 42.34)$ phonons, $k_B$ standing for Boltzmann's constant. 

The experiment setting described above is well captured by the one dimensional motion of an oscillator undergoing dissipation and fluctuations. We model the dynamics of the position $z(t)$ of the oscillator along the relevant direction as
\begin{equation}    \label{Langevin_1D}
    \ddot{z}(t) = -\Gamma_m \dot{z}(t) + S(t)\frac{F(z)}{m} + \frac{\mathcal{F}_\text{fluct}(t)}{m}, 
\end{equation}
where $\Gamma_m$ is the damping rate induced by the effects of the residual background gas in the trap. Such rate can be quantified as~\cite{gieseler2013thermal}
\begin{equation}    \label{gas_damping_Gieseler}
    \Gamma_m = \frac{64Pr^2}{m\overline{v}_\text{gas}},
\end{equation}
with $P$,  $T$, $\bar v_\text{gas}=\sqrt{k_B T/m_\text{gas}}$ being the pressure of the gas,  its temperature, the root mean square velocity of its constituents and the particle mass, respectively.
The second term in the right-hand-side of Eq.~\eqref{Langevin_1D} describes the restoring force originating from the interaction 
with a Gaussian light beam in the Rayleigh regime.  
This can be computed from the profile of the intensity of the trapping laser beam $I$, namely one has
\begin{equation}    \label{gradient_force}
    F(z) = \frac{2\pi n_{m}r^3}{c} \left(\frac{n_{r}^2-1}{n_{r}^2+2}\right)\frac{\partial I}{\partial z} ,
\end{equation}
with $n_{r}=n_{p}/n_{m}$, $n_{m}$ and $n_{p}$ being the refractive index of the medium and of the silica particle (here assumed to take the nominal values of 1 and 1.44, respectively), and $c$ the speed of light.
In the large oscillation-amplitude limit one needs to account for the non-linearities of the trapping potential and thus $F(z)$ takes a non-linear position dependence $F(z)=-m \omega_z^2z(1+\zeta z^2)$, with $\zeta = -1.107 \, (\mu\text{m})^{-2}$ \cite{flajvsmanova2020using, setter2019characterization}. For sufficiently small amplitudes $|z|\ll 30.16 \, \text{nm}$, one can claim for the harmonic approximation and neglect higher order terms. The control protocol is imprinted via the modulation function $S(t)$, which reflects the amplitude variation of the optical field.
The last term in Eq.~\eqref{Langevin_1D} accounts for the stochastic fluctuations due to the collisions between the nanoparticles and the  residual gas. A standard statistical mechanics approach leads to the expression~\cite{kubo1966fluctuation}
$\mathcal{F}_\text{fluct}(t)=\sqrt{2\Gamma_{m}k_{B}T}\eta(t)$, where  $\eta(t)$ is the zero-mean Gaussian white noise with two-time correlator $\langle \eta (t) \eta (t') \rangle = \delta(t - t')$.
Additionally, one can add the following acceleration term
\begin{equation}
    \frac{F_{fb}(z)}{m}= -\gamma_{fb}\sin{\Big(\arctan{\frac{v_{z}}{z\omega_{z}}}\Big)},
\end{equation}
to account for the feedback cooling of the particle. Importantly, the feedback acts on the particle only if the latter is sufficiently localised and if the feedback is locked to its motion.

\noindent
{\it Description and performance of the protocol}  - The initially cooled state is expanded by non-adiabatic switching between different trap frequencies~\cite{rashid2016experimental}, as  proposed in Ref.~\cite{janszky1986squeezing,wu2024squeezing}. Gradual and steady expansion is achieved by concatenating $1000$ pulses that first lower the laser power by $10\%$ for $\tau_{low}=\pi/2\omega_{z}\sqrt{S} =\SI{3.4}{\mu s}$ 
and  and then raise the power back up to its initial value for $\tau_{high}=\pi/2\omega_{z} = \SI{3.22}{\mu s}$ as shown in Fig.~\ref{fig:wide_panel}B. Feedback cooling is kept on throughout the whole protocol but becomes ineffective after $t=0$, when the pulses begin. A third order band pass filter with $\SI{14}{kHz}$ bandwidth around $\omega_{z}/2\pi$ is applied to the signal to eliminate contributions from the $x$ and $y$ directions as well as the modulation of control signal $\omega_{S}/2\pi=1/(\tau_{low}+\tau_{high})=1.95(\omega_{z}/2\pi)$. The protocol is repeated $671$ times, yielding an ensemble of trajectories that are synchronized with respect to the expansion pulses. The phase space probability density distribution associated with the initial cold state of the particle is circular but the pulses transforms into ellipsoidal bivariate Gaussians as seen in Fig.~\ref{fig:wide_panel}C.

The distribution associated with the initial cooled state of the particle is isotropic. A sudden decrease in trap stiffness reduces the potential energy while leaving the kinetic energy unchanged. During the subsequent quarter of the oscillator period, the different quadrants of the density function evolve distinctly. Points along the position axis, having lost potential energy, rotate toward the momentum axis as they convert into kinetic energy, reaching smaller momentum values. Conversely, points along the momentum axis retain their kinetic energy and evolve toward extended positions in the softer trap. This process effectively transforms the circular phase space distribution into an ellipsoidal shape.

The rising pulse returns the trap frequency to its original value and the ellipse is captured with its semi-major axis, $\sigma_{max}$, extended along the position axis. During the next quarter of the oscillator period, the dynamics in the stiffer trap project $\sigma_{max}$ further along the momentum axis, while the already small momentum values recede to even lower values in the position axis.

\begin{figure*}
\centering
\includegraphics[scale=1]{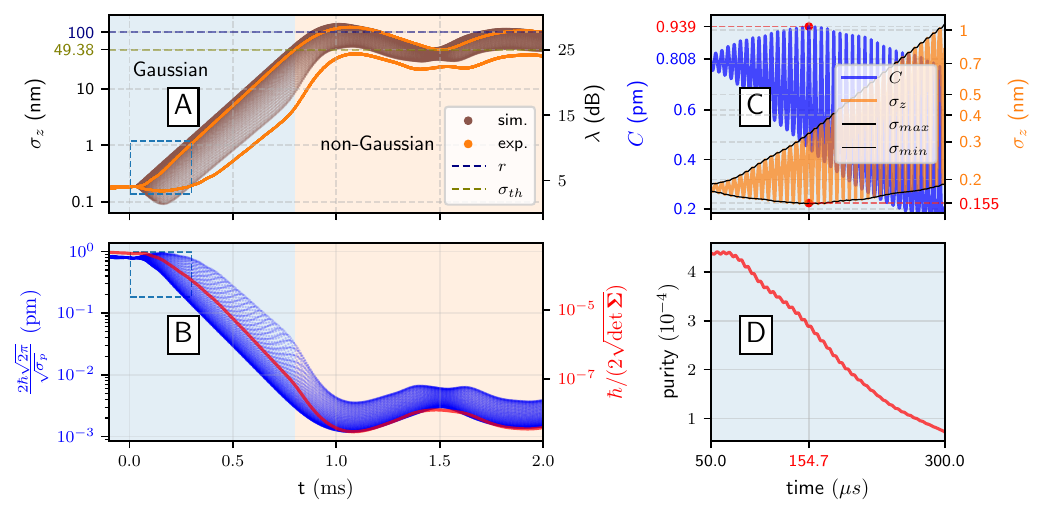}
\caption{\label{fig:experiment_coherence_expansion}{\bf } {\bf Panel A}: Time-evolution of the position standard-deviation $\sigma_{z}$, experiment vs simulations. The length of the semi-major axis $\sigma_{max}$ increases exponentially in a timescale of about \SI{1}{ms} all the way up to \SI{124}{nm}, which is of the order of the thermal spread (green dashed line) and radius of the nanoparticle used in the experiment (blue dashed line). In the same time-window, the semi-minor axis undergoes an initial contraction,  as one would expect from the dynamics. However, such contraction is then followed by an expansion occurring at roughly the same rate as the growth of the semi-major axis, suggesting additional heating mechanisms. The brown dots are the results of our Langevin simulations where we added a Gaussian noise to the harmonic frequency $\omega_z$. {\bf Panel B}: State Coherence $C(\hat \rho)$ (blue trace) and purity $P(\hat \rho)$ (red line) evolutions being computed as in Eq.~\eqref{eq.coherence.gauss} and Eq.~\eqref{eq.purity.gauss} respectively. The state's coherence initially increases, up to $154.7 \, \mu s$ when it peaks at \SI{0.939}{pm}. This point corresponds to the semi minor axis at its smallest value $0.155 \,\text{nm}$. Afterwards the state becomes broader, and coherence starts to decrease. After an initial plateau, the purity decreases exponentially. After around 1\,ms, it follows a similar oscillatory behavior as coherence. {\bf Panel C}: Comparison of coherence (blue line) and position standard-deviation (orange line) at the beginning of the protocol. We highlighted the maximal value of coherence, which is larger than the initial one, and the minimal value of the position standard-deviation. {\bf Panel D}: Purity evolution at the beginning of the protocol, where the initial plateau is clearly visible.} 
\end{figure*}

The state expansion is captured by the behavior of the semi-major axi, $\sigma_{max}$ illustrated in Fig.~\ref{fig:experiment_coherence_expansion}A. Within the first \SI{0.8}{ms} of the protocol, the system remains in the harmonic approximation of the potential and the expansion is exponential, characterized by a time-constant of $\tau_\text{protocol} = 10\times (\tau_{low}+\tau_{high}) =  \SI{68.6}{\mu s}$ that is much faster than the thermalization process due to gas collisions. The latter occurs within a timescale of seconds at the experimental pressure, as described by Eq.~\eqref{gas_damping_Gieseler}. After the initial exponential expansion, the non-linearity of the trap becomes relevant and the expansion starts a damped oscillatory behavior, stabilizing motion into a non-equilibrium steady-state configuration, characterized by a non-Gaussian multi-modal distribution~\cite{muffato2024generation}. After \SI{6.7}{ms}, the expansion pulses stop to leave room for feedback, which causes the non-equilibrium steady-state to relax toward the cold state within a typical time of $\tau_\text{feedback}= \SI{44}{ms}$. 

{The semi-minor axis $\sigma_{min}$ shortly decreases  within the first $~155 \, \mu s$, after which a broadening trend begins. This is an interesting feature: Ideally, one would expect for $\sigma_{min}$ to decrease at the same rate as the expansion of the semi-major axis. This is clearly not the case. In order to gain insight into such feature, we have performed extensive numerical simulation of the Langevin dynamics of the motional degrees of freedom with added Gaussian noise on the frequency of the oscillatory motion, averaging over a statistical ensemble of runs. The results of our simulations reproduce very closely the main features of the experimental data, allowing us to conjecture that laser intensity fluctuations cause small changes in $\omega_{z}$, thus mutually offsetting each position trajectory and resulting in a broadening of $\sigma_{min}$.}

Having illustrated the phase-space effects of the control scheme that we have put in place, we now  quantify the degree of state expansion. In Fig.~\ref{fig:experiment_coherence_expansion}A the standard deviation of the position quadrature at a generic time during the protocol is compared to its starting value. The (in general) elliptical phase-space distribution rotates due to the dynamics of the particle. 
This reflects in the broadening of the curve, as the single oscillations cannot be seen due to their timescale $\sim\pi/\omega_z=6 \times 10^{-6}\,$s.
The initial exponential expansion continues up to $t_{peak}=1.07$\,ms, when we reach the first peak of the curve. At such time, we have an expansion of $\lambda=10\log_{10}(\sigma_z(t_{peak})/\sigma_\text{cold})=\SI{28.4}{dB}$, which corresponds to $\sigma_z(t_{peak})=\SI{124}{nm}$. Remarkably, this is 24\% larger than the nanoparticle dimension, and 158\% larger than the corresponding thermal spread at \SI{300}{K}, which is quantified by the expression~\cite{feynman2010quantum}
\begin{equation}
\sigma_\text{th}=\sqrt{\frac{\hbar}{2m \omega_z}\coth\left(\frac{\hbar \omega_z}{2k_B T}\right)}.
\end{equation}

\noindent
{\it Purity and coherence} -- We proceed to certify that the expansion is {\it coherent} and not a by-product of environmental-induced diffusion or other spurious effects. In particular, we compute the purity of the motional state of the particle  
\begin{equation}\label{eq.purity.cv}
  P(\hat \rho)=\operatorname{Tr}[\hat \rho^2]=\int\D x\braket{x|\hat \rho^2|x}.
\end{equation}
For our experiment, the quadratic approximation of the potential is valid up to 800\,$\mu$s. From such moment on, the wavefunction explores also the non-quadratic part of the potential. Within such timeframe, the motional state of the particle is approximately Gaussian and thus fully described by the associated covariance matrix $\bm{\Sigma}(t)$ whose entries read $\Sigma_{ij}(t)=\mathrm{Cov}(r_i(t),r_j(t))$ with $\bm r=(z,p_z)$. In such a case, the purity takes the following form
\begin{equation}\label{eq.purity.gauss}
    P(\hat \rho)=\frac{\hbar}{2\sqrt{\det{\bm \Sigma}}}.
\end{equation}
Purity gives a direct indication of the unitarity of the dynamics of the system, as it remains constant if the dynamics is unitary and decreases under decoherence (e.g.~a thermalisation process) and increases under purification protocols (e.g.~a energy decay leads a statistical mixture in the ground state, which is a pure state). Purity, however, does not provide indications if  our protocol maintains or enhances position coherence of the system, which instead can be quantified by a coherence measurement in position basis. Although the latter is not well defined in continuous variables systems \cite{streltsov2017colloquium}, we can construct a suitable version of the discrete $l_1$-coherence measure $C_{l_1}=\sum_{x_i\neq x_j}|\braket{x_i|\hat \rho|x_j}|$ as
\begin{equation}\label{eq.coherence.cv}
     C(\hat \rho)=\int\D x\int\D y\,|\braket{x|\hat \rho|y}|,
\end{equation}
which has the dimensions of a length.
Conceptually, the difference between $C_{l_1}$ and $C(\hat \rho)$ is given by the sum of the terms on the diagonal ($x_i=x_j$), which are missing in the former and accounted for by the latter. As the sum of diagonal terms equals the trace of the state, this contribution provides a unit off-shift that we wash out by considering the coherence variation $\Delta C=C(\hat \rho_\tau)-C(\hat \rho_0)$ between the times $t=\tau$ and $t=0$. For a Gaussian state with covariance matrix ${\bm \Sigma}$, we have 
\begin{equation}\label{eq.coherence.gauss}
    C(\hat \rho)=\frac{2\hbar\sqrt{2\pi}}{\sqrt{\sigma_p}},
\end{equation}
where $\sigma_p$ is the momentum variance. 
For comparison, the purity and the coherence measure for a thermal state of thermal occupation number $\bar{n}$ read
\begin{equation}
P(\hat\rho_\text{th})=\frac{1}{2\bar n+1},
\quad \text{and}\quad 
    C(\hat \rho_\text{th})=\frac{4\sqrt{\hbar \pi}}{\sqrt{m \omega_z(2\bar n+1)}}
\end{equation}
respectively. The evolution of coherence and purity are shown in Fig.~\ref{fig:experiment_coherence_expansion}B, C and D. {The former oscillates between $C_{min}$ and $C_{max}$ at the same frequency of $\sigma_z$. Its maximum value increases from 0.808 to 0.939, achieved at $t=154.7\,\mu$s, showing that the particle undergoes a coherent-expansion dynamics. Conversely, after a short plateau, the state purity decreases  exponentially with the same characteristic time of the protocol's expansion $\tau_\text{protocol}= \SI{68.6}{\mu s}$.} Notably, our coherence measure is proportional to the coherence length $l=\sqrt{8\sigma_x}P(\hat \rho)=C(\hat \rho)/\sqrt{4\pi}$ employed in Refs.~\cite{rossi2024quantum,mattana2025trap}. In terms of the latter figure of merit, our protocol achieves a relative variation of coherence on $\sim 16\%$, which is remarkable considering the large thermal occupation number and low purity of our initial state. For comparison, a  larger coherence variation $\sim 250\%$ was obtained in \cite{rossi2024quantum} with a similar experimental setup, but with a much more modest position expansion of $\sim 8$\,dB. Conversely, in \cite{mattana2025trap} a 22.7\,dB expansion was achieved with a free-fall protocol, but with a coherence variation amounting to $\sim-87\%$. 

\noindent
{\it Conclusion --} We have shown experimentally that the pre-cooled motional state of a levitated optomechanical system can be expanded coherently. Our results show that the resulting motional state can exceed the physical dimensions of the nanoparticle itself, reaching a remarkable degree of {spatial expansion} of nearly 30\,dB. {Moreover, such degree of coherence can be preserved during the protocol despite the strong presence, in our setting, of environmental and laser-intensity noise. Improvements the stability of the latter parameter will allow for higher coherence figures at larger expanded states, thus} opening up a fruitful route in the quest for exploring the wave-like nature of massive quantum states. In turn, higher degree of coherent state expansion -- which we will achieve by preparing the motional state of the particle at lower effective temperatures -- will unlock the use of levitated nanoparticles in quantum sensing applications, where the enhanced spatial coherence will be crucial to achieve unprecedented degrees of  sensitivity and precision.

\noindent
{\it Acknowledgements --} 
We acknowledge discussions with Diana Chisholm, Elliot Simcox, Jack Homans, Chris Timberlake, Jakub Wardak and Qiongyuan Wu. We acknowledge funding from the EPSRC Network Grant {Levinet} (EP/W02683X/1). We further acknowledge financial support from the UK funding agency EPSRC (grants EP/W007444/1, EP/V035975/1,  EP/V000624/1, EP/X009491/1 and EP/T028424/1), the Leverhulme Trust (RPG-2022-57), the Royal Society Wolfson Fellowship (RSWF/R3/183013), the Department for the Economy of Northern Ireland under the US-Ireland R\&D Partnership Programme, the PNRR PE Italian National Quantum Science and Technology Institute (PE0000023), the University of Trieste (Microgrant LR 2/2011) and the EU Horizon Europe EIC Pathfinder project QuCoM (GA no.~10032223). 

\noindent
{\it Data availability statement --} Data sets are available at server: https://doi.org/10.5258/SOTON/D3673, and analysis codes is available from the authors upon reasonable requests.

\providecommand{\noopsort}[1]{}\providecommand{\singleletter}[1]{#1}%

\end{document}